\documentclass[epjST]{svjour}
\usepackage{graphics,graphicx}
\usepackage{amsfonts}
\usepackage{amsmath}
\usepackage{amssymb}
\begin{document}
\title{Ratchet transport with
subdiffusion}
\author{S. Denisov \thanks{\email{sergey.denisov@physics.uni-augsburg.de}} }
\institute{Institute of Physics, University of Augsburg, D-86135
Augsburg, Germany}
\abstract{We introduce a model which incorporate the subdiffusive
dynamics and the ratchet effect. Using a subordination ideology, we
show that the resulting directed transport is sublinear, $\langle
x(t)\rangle \simeq Jt^{\beta}$, ~$\beta < 1$. The proposed model may
be relevant to a phenomenon of saltatory microbiological motility.
} 
\maketitle

Thermal fluctuations alone cannot create a steady transport in an
unbiased system. Luckily, microbiological realm operates far from
equilibrium \cite{non}, where directed motion can appear under
nonequilibrium conditions \cite{Ratchet1}. The corresponding
\textit{ratchet} effect  has been proposed as a physical mechanism
of a microbiological motility \cite{Ratchet2}. The nonequilibrium
conditions might induce  another intriguing peculiarity of
microbiological transport; namely, the \textit{anomalous} diffusion
\cite{Anomalous1}. The quasi-random wandering at the molecular scale
can be characterized by a mean square displacement (msd),
$\sigma(t)=\langle x^{2}(t)\rangle - \langle x(t)\rangle^{2}$,
which, in many cases, follows a power law, $\sigma(t) \sim
t^{\alpha}$, $\alpha \neq 1$, rather than the linear time dependence
for a Brownian particle \cite{Biolog1,Golding}. $\alpha > 1$
corresponds to enchance or superdiffusion, while $\alpha < 1$ to
subdiffusion \cite{Anomalous1,Klafter_Rep}.

Directed current and diffusion are generated by the same trajectory,
$x(t)$, and  characterized by the first and the second moments of
the same process being therefore strongly conjugated. In the case of
the normal diffusion and the superdiffusion, due to finiteness of
all statistical moments, a directed transport, if any, corresponds
to a linear grow of the mean displacement, $\langle x(t) \rangle
\sim t$, with the corresponding current $J=\lim_{t\rightarrow
\infty} {x(t)}/t$ \cite{our_ratchet}.

For the subdiffusion a situation is less obvious. In this regime the
motion is made up of periods of sticking events separated by fast
jumps to a new position (see, f.e., Ref. \cite{Golding}). Formally,
the broad power-law distribution of sticking times,
\begin{equation}
\psi(t) \sim t^{-1-\beta},~~~ 0 < \beta < 1, \label{Levy}
\end{equation}
leads to the subdiffusion with an exponent $\alpha=\beta$
\cite{Klafter_Rep}. The \textit{saltatory} molecular motor's
transport \cite{saltatory}, rapid bursts of directed movements
interrupted by pauses of variable duration, has been tracked within
a cell  by using microinjected fluorescent beads \cite{bead}. The
abovementioned observations call for a study of a ratchet transport
mediated by anomalous trapping events.

In this paper we propose a simple model which naturally incorporates
both mechanisms, subdiffusion and ratchet effect, thus bridging two
research lines that were so far basically disconnected from one
another. Using subordinated ideology \cite{Sokolov} we show that the
subdiffusion in a ratchet potential results in the sublinear
directed transport, $<x(t)> \simeq Jt^{\beta}$. The subordinate
formalism enables us to reformulate the problem within the circle
map's theory and to derive  necessary and sufficient conditions for
the directed current appearance.

We start with the model which describes a dynamics of the overdamped
particle exposed to the shot-noise,
\begin{equation}
\dot{x}=\sum a_{j}(x)\delta(t-t_{j})  \label{model}
\end{equation}
where $a_{j}$ gives the length and the direction of the
corresponding step taking place at the time instant $t_{j}$. We
assume that the length $a_{j}$ depends locally on a periodic
potential $U(x)$, $U(x+L)=U(x)$, and a noise $\xi$, such that
\begin{equation}
a_{j}=-U'(x(t_{j}))+\xi(t_{j}). \label{coeff}
\end{equation}
The model (\ref{model}-\ref{coeff}) can be treated as the overdamped
limit of the standard model \cite{ott}. The corresponding process,
$x(t)$, can be considered not as a function of time $t$, but rather
as a function of the number of steps, $n$. The dynamics of the
system (\ref{model}) can be represented as the noised circle map
\cite{ott},
\begin{eqnarray}
x_{n+1}=x_{j}+f(x_{n})+\xi_{n}, \label{map}
\end{eqnarray}
where  $f(x)=-U'(x)$ stands for the acting force. The process $x(t)$
is \textit{subordinated} \cite{Sokolov} to the map (\ref{map}), such
that the time is governed by the linear map,
\begin{eqnarray}
t_{n+1}=t_{n}+ \triangle t_{n}.
\end{eqnarray}

In addition, here we assume that \textit{(i)} the dispersion is
finite, $\langle a_{j}^{2} \rangle < \infty$, and \textit{(ii)} the
time between steps, $\triangle t_{j}=t_{j+1}-t_{j}$, is a random
stationary process with the probability density function (pdf)
$\psi(\triangle t)$. If $\psi(\triangle t)$ has a finite first
moment, $\langle \triangle t \rangle < \infty$, and $a_{j}$ is
independent on $x$ and has a symmetrical distribution (Gaussian,
Poisonian, etc) then for the time scale $t \gg \langle \triangle t
\rangle $ we get normal Gaussian diffusion, $<x^{2}(t)> \sim t$,
which can be described  by an ordinary Langevin equation
\cite{Sokolov}. If $\psi(t)$ has a divergent first moment, which is
the case of the distribution (\ref{Levy}), then we deal with the
subdiffusion, where the corresponding  msd's exponent is
$\alpha=\beta$ \cite{Klafter_Rep}.

The role of the sticking time reduces to the fact that the actual
number of steps made up to the time instant $t$ fluctuates, so the
operational time $n$ is a random function of the physical time $t$.
This function, however, is monotonously nondecaying with $t$ and
thus allows a causal ordering of the events. For the pdf $p(x,t)$
one has \cite{Sokolov}
\begin{eqnarray}
p(x,t)=\sum_{n}W(x,n)\chi_{n}(t),
\end{eqnarray}
where $W(x,n)$ is the pdf for the iterated process (\ref{map}), and
$\chi_{n}(t)$ is the probability to make exactly $n$ steps up to
time $t$.

The asymptotic transport is $\langle x(t)\rangle = J \cdot \langle
n(t) \rangle$, where $\langle n(t) \rangle=\sum_{n=0}^{\infty} n
\chi_{n}(t)$, and the current value $J$ follows from transport
properties of the map (\ref{map}), $\langle x(n)\rangle \simeq Jn$.
By using the Laplace transform in the time domain, it can be shown
that $\langle \tilde{n}(s)\rangle = \tilde{\psi}/s(1-\tilde{\psi})$,
where $\psi(t)$ is the pdf for sticking time. For the Poissonian
process, $\tilde{\psi}(s)=\nu/(s+\nu)$, one can easily get  $\langle
n(t) \rangle=\nu t$. For the power-law pdf $\psi(t)$ follows that
$\langle \tilde{n}(s) \rangle \approx \tau^{-\beta}s^{-1-\beta}$
and, finally, $\langle n(t) \rangle \approx
\frac{t^{\beta}}{\tau^{\beta}\Gamma(1+\beta})$, so that
\begin{equation}
\langle x(t)\rangle \approx \frac{J}{\tau^{\beta}\Gamma(1+\beta)}
t^{\beta}, \label{eq:current}
\end{equation}
where $\Gamma(x)$ is the Gamma function.

The subordination approach allows us to separate  transport
properties, precisely the value of the generalized current $J$,
which follows from the map (\ref{map}), from the sublinear
asymptotic (\ref{eq:current}), which is governed by the sticking
time pdf $\psi(t)$.

As an illustrative example we consider here the two-harmonics
potential force,
\begin{equation}
f(x)=E_{1}\sin(2\pi x)+E_{2}\sin(4\pi x+\theta), \label{climb}
\end{equation}
which transforms the system (\ref{map}) into a ratchet version of a
climbing circle map \cite{ott}. Without loss of generality, we chose
here the Gaussian noise $\xi$ with the dispersion $\eta^{2}$.

\begin{figure}[t]
\includegraphics[width=0.7\linewidth,angle=0]{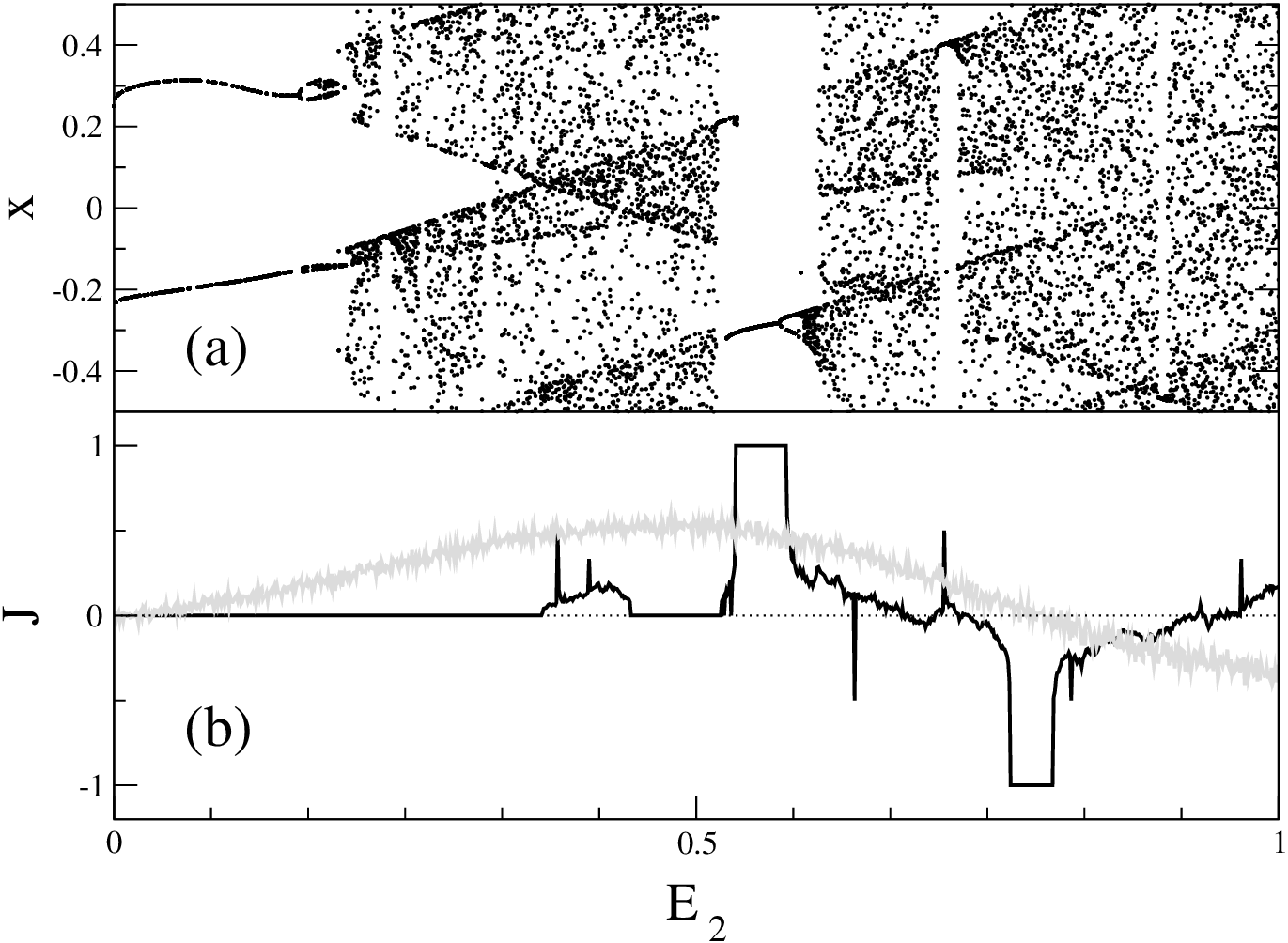}
\caption{(a) The bifurcation diagram and (b) the  current value $J$
as functions of $E_{2}$ for the map (\ref{map}-\ref{climb})
($E_{1}=0.5$). Black line corresponds to deterministic case,
$\eta^{2}=0$,  and the grey line corresponds to noised case,
$\eta^{2}=0.1$. Note that the current value in the last case is
scaled by the factor $40$.} \label{fig:p1}
\end{figure}

Certain symmetries of the Eqs.(\ref{model}, \ref{climb}) need to be
broken in order to fulfill necessary conditions for a directed
transport appearance  \cite{flach}. Suppose that there is a
transformation which leaves the equation (\ref{model}) invariant and
changes either the sign of $x$, $x\rightarrow -x$, or invert time,
$t \rightarrow -t$ (but not both operations simultaneously!). Such a
transformation  maps a given trajectory into another one, with the
opposite velocity. If at least one of such symmetries exists, then
contributions of the trajectory and its symmetry-related counterpart
will cancel each other and the asymptotic current $J$ will be equal
to zero \cite{flach}.

In order to fulfill the necessary condition for the dc-current
appearance, the following symmetries of the potential force should
be violated:
\begin{eqnarray}
f(x)=-f(-x)~~(\textrm{or}~~U(x)=U(-x)), \\
f(x)=-f(x+L/2)~~ (\textrm{or}~~U(x)=-U(x+L/2)) \label{symm},
\end{eqnarray}
which are the reflection- and the shift-symmetry following to
Ref.\cite{flach} and termed as "symmetry" and "supersymmetry" in
Ref.\cite{Reim2}. Both the symmetries are violated when $E_{2}\neq
0$ and $\theta \neq k\pi$.

The symmetry violation is the necessary condition for the current
appearance. The  current  value is determined by microscopic
dynamical mechanisms \cite{denisov_flach}. It is reasonable then to
start the analysis of  map's transport properties from the
deterministic limit, $\xi=0$. Fig.1 shows the bifurcation diagram
and the current value as  functions of the second harmonic
amplitude, $E_{2}$. There is an evident relationship between
multiple current reversals and different kinds of bifurcations,
which is a general property of \textit{underdamped} deterministic
ratchets \cite{Mateos}. Here, at the \textit{overdamped} limit, this
relationship can be explained qualitatively. The current value is
equal to the average force,
\begin{equation}
J=\langle f(x) \rangle = \int_{0}^{L}dx f(x)\hat{P}(x),
\label{current}
\end{equation}
where $\hat{P}(x)=\sum_{n=-\infty}^{\infty}P(x+nL)$, $n \in
\mathbb{Z}$, is the reduced invariant density for the map
(\ref{map}-\ref{climb}). Since any critical bifurcation, like a
tangent bifurcation (transition from a limit cycle to a chaotic
regime) \cite{ott}, always accompanied by drastic changes of the
invariant density $P(x)$, such a bifurcation leads to a  "jump" of
the current value.

The addition of a  noise changes the rectification dynamics, but
still the system generates a non-zero current (gray curve in
Fig.1b). However, the invariant density even at a weak noise limit
converges to the Boltzmann pdf \cite{hanggi_map},
$\hat{P}(x)=\frac{1}{N} \exp (-U(x)/\sigma^{2})$, so that the
integral in rhs of Eq.(\ref{current}) goes to zero. Thus, the
current decays rapidly with the increasing of the noise strength
(Fig.1b).

\begin{figure}[t]
\includegraphics[width=0.75\linewidth,angle=0]{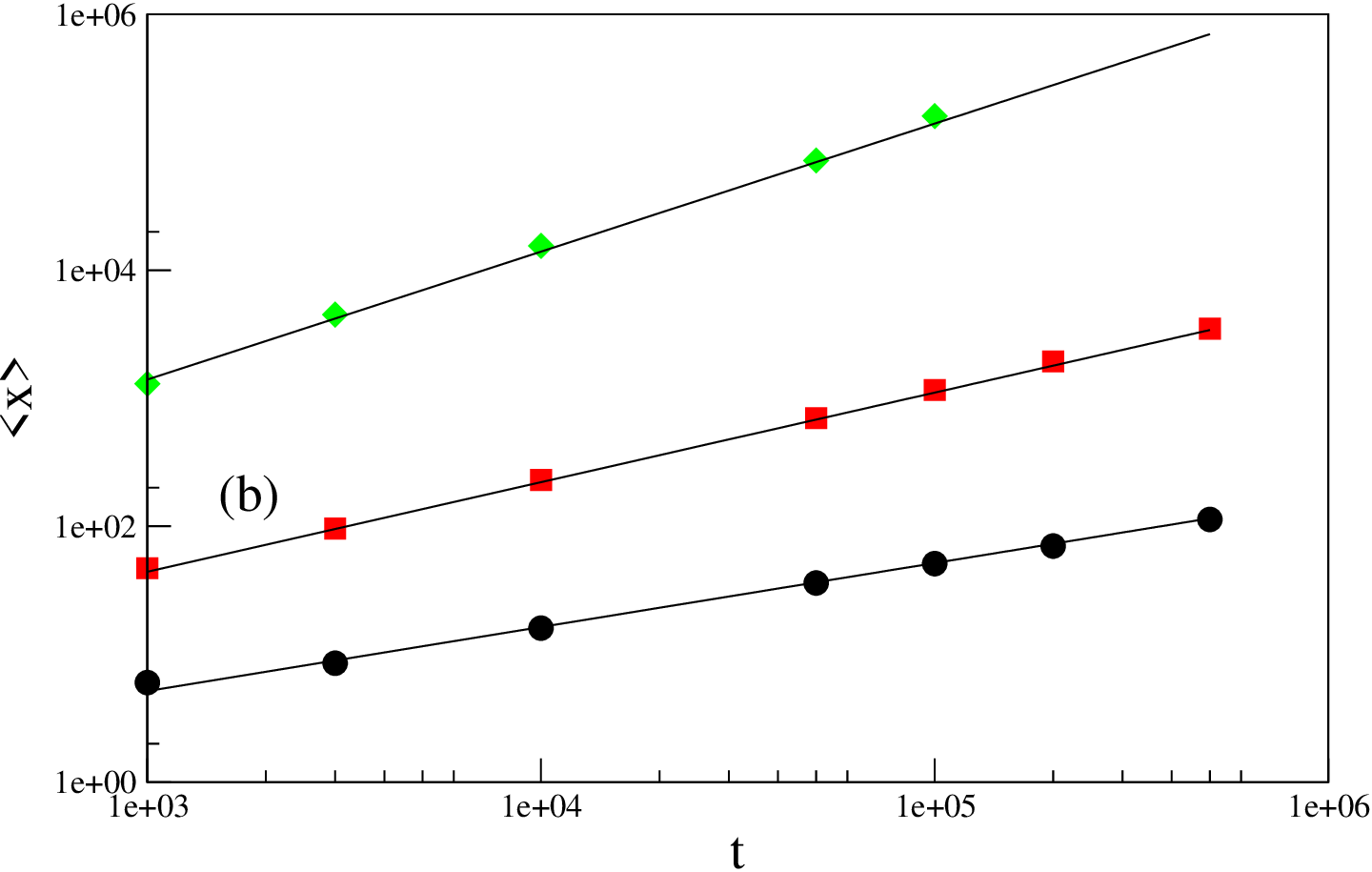}
\caption{ The evolution of the ensemble mean displacement $\langle
x(t) \rangle$ versus $t$ for the different values of subdiffusive
exponent: $\alpha=0.5$ ($\bullet$), $\alpha=0.7$ ($\blacksquare$),
and $\alpha=1.2$ ($\blacklozenge$). Lines correspond to asymptotics
from Eq. (\ref{subcurrent}). The parameters are $N=10^{5}$,
$E_{1}=0.5$, $E_{2}=0.8$, $\theta=-2.24\pi$, and $\eta^{2}=0.02$.}
\label{fig:p2}
\end{figure}

We furthermore assume that a time between subsequent kicking events
is a random stationary process with the  pdf (\ref{Levy})
\footnote{For the generation of the random variable $\triangle t$
with the pdf (\ref{Levy}) we have used the random variable $\xi$
with the uniform distribution on the unit interval, $[0,1]$, and the
transformation $\triangle t=\triangle t_{c}\xi^{-1/\beta}$.}.
 The asymptotic displacement can be written as
$\langle x(t)\rangle = J \cdot \langle N(t) \rangle$, where the
current value $J$ follows from transport properties of the map
(\ref{map}-\ref{climb}), $\langle x(n)\rangle \simeq Jn$. Formally,
we get
\begin{equation}
\langle x(t) \rangle \simeq Jt^{\alpha}. \label{subcurrent}
\end{equation}

We  consider now a large ensemble of noninteracting ratchets. The
dynamics of each particle in the operational time frame is governed
by the same map (\ref{map}-\ref{climb}), but the physical time $t$
is different for different particles. Nevertheless, the
subordination formalism allows to make casual ordering of events
\cite{Sokolov}. Thus we can calculate the displacement $\langle x(t)
\rangle$ by using an ensemble averaging. In Fig.2 we shown the
evolution of the mean displacement, $\langle x(t)\rangle$ , for
different values of the waiting time exponent $\alpha$.

Spatial distributions for different times are shown on Fig.3. For
the case of normal diffusion (Fig.3a), $\alpha > 1$, the evolution
follows an universal Gaussian scaling, which has been found for
weakly \textit{underdamped}   ratchets \cite{ratchet_Hanggi}. This
Galilei invariant Brownian process, $p(x,t)\simeq
\frac{1}{\sqrt{t}}g(\frac{x-Jt}{\sqrt{t}})$, where $g(x)$ stands for
the Gaussian pdf, is very different from the Galilei variant
subdiffusive ratchet regime \cite{Klafter_Rep}. The corresponding
pdf is asymmetric with respect to its cusp-like maximum which stays
fixed at the origin, and the plume stretches more and more into
transport direction (Fig.3b). This behavior is reminiscent of a
subdiffusive dispersive transport under a constant tilting force
\cite{Mon2}.

\begin{figure}[t]
\includegraphics[width=0.8\linewidth,angle=0]{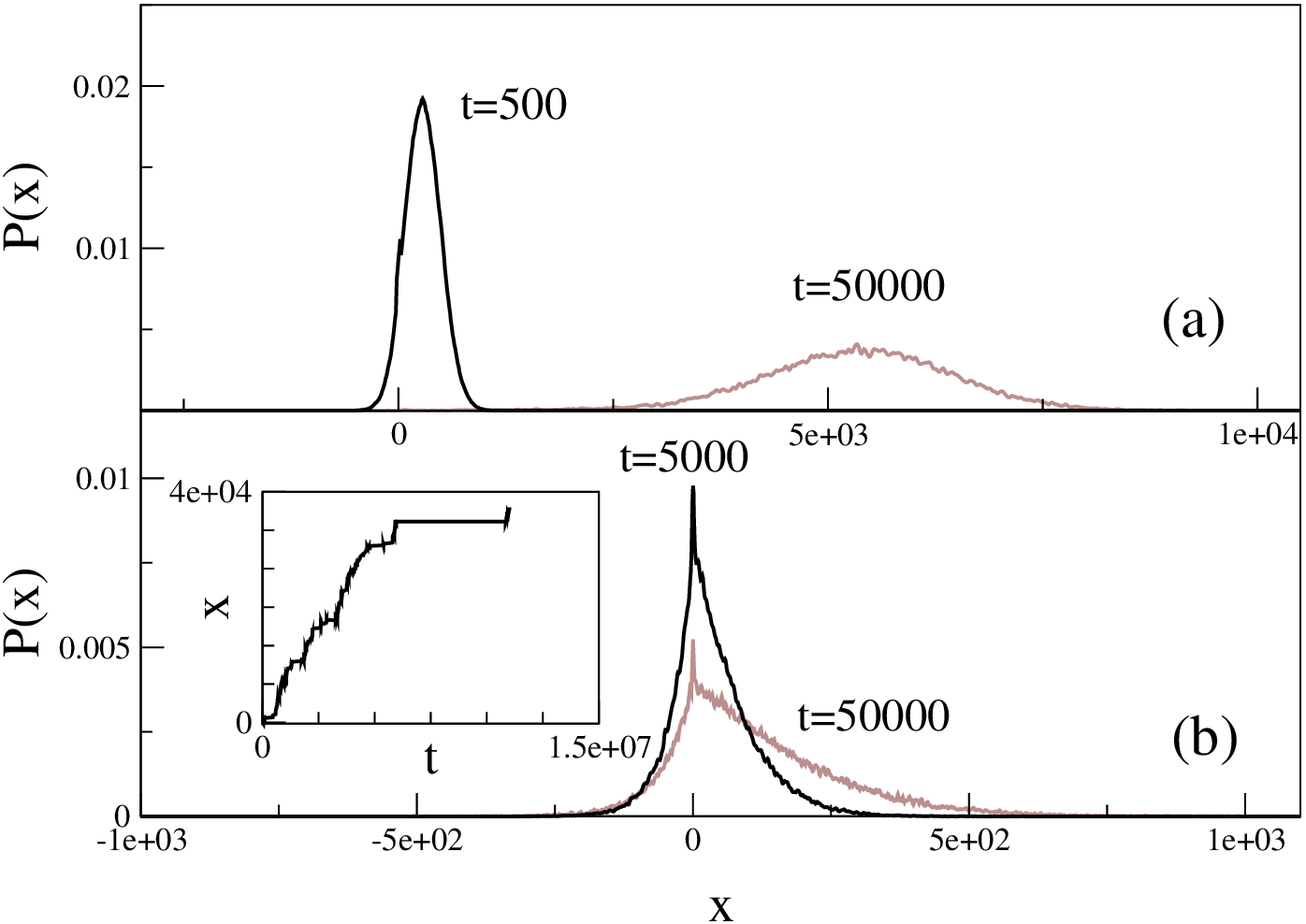}
\caption{Spatial distributions for the ensemble of  $N=10^{6}$
single ratchets for (a) $\alpha=1.2$  and (b) $\alpha=0.5$ for
different times $t$. The inset shows single particle's trajectory.
Other parameters are as in Fig.2.} \label{fig:conv}
\end{figure}

Summing up, we have considered the new model which yields anomalous
ratchet dynamics. Transport properties, such as direction and value
of the generalized current stem from the ratchet-like periodic
potential. The anomalous character of a kinetics is governed by the
waiting time pdf. The proposed model provides further contribution
to the studies of the microbiological transport. It sets up the link
between ratchets and a power-stroke approach to a microbiological
transport, still existing dichotomy \cite{dichotomy}. The model can
also be useful for an analysis of experimentally detected saltatory
motility in a cell by using of microinjected fluorescent beads
\cite{bead}.


\begin{thebibliography}{}
\bibitem{Ito} The appearance of the Ito's form is consistent with the map
in Eq.(15).
\end{thebibliography}
\end{document}